\documentclass[aps,prl,twocolumn,superscriptaddress,amsmath,floatfix]{revtex4-2}

\usepackage{graphicx}
\usepackage{dcolumn}
\usepackage{bm}
\usepackage{amssymb}
\usepackage{hyperref}
\usepackage{multirow}
\usepackage{color}

\begin{document}

\title{Phase separation in a chiral active fluid of inertial self-spinning disks}

\author{Pasquale Digregorio}
\email{lino.digregorio@gmail.com}
\affiliation{Dipartimento di Fisica, Universit\`a degli studi di Bari and INFN, Sezione di Bari, via Amendola 173, I-70126, Bari, Italy}
\affiliation{Universitat de Barcelona Institute of Complex Systems (UBICS), Universitat de Barcelona, 08028 Barcelona, Spain}
\author{Ignacio Pagonabarraga}
\email{ipagonabarraga@ub.edu}
\affiliation{Universitat de Barcelona Institute of Complex Systems (UBICS), Universitat de Barcelona, 08028 Barcelona, Spain}
\affiliation{Departament de F\'{\i}sica de la Mat\`eria Condensada, Universitat de Barcelona, Carrer de Mart\'{\i} i Franqu\'es 1, 08028 Barcelona, Spain}
\author{Francisco Vega Reyes}
\email{fvega@eaphysics.xyz}
\affiliation{Departamento de F\'{\i}sica and Instituto de Computaci\'on Científica Avanzada (ICCAEx), Universidad de Extremadura, 06071 Badajoz, Spain}

\date{\today}

\begin{abstract}
We show that  systematic particle rotations in a fluid composed of disk-shaped spinners can spontaneously lead to phase separation. 
The phenomenon arises out of a homogeneous and hydrostatic stationary state, due to a pressure feedback mechanism that increases local density fluctuations. 
We show how this mechanism induces phase separation, coined as Rotation Induced Phase Separation (RIPS), when the active rotation is not properly counterbalanced by translational friction. 
A low density phase can coexist with a dense chiral liquid due to the imbalance between pressure and stress transmitted through chiral flows when a significant momentum transfer between rotational and translational motion can be sustained. 
As a consequence, RIPS is expected to appear generically in chiral fluids.
\end{abstract}

\maketitle

Intense and active research on complex fluids dynamics during the last decades has expanded the limits of non-equilibrium statistical mechanics and hydrodynamics, leading to generalized theories that describe a wide variety of non-equilibrium systems~\cite{Tsimring2006,Volpe2016}. 
The need for these generalized theories stems from strong experimental and computational evidence that cannot be explained within the context of classical fluid and statistical mechanics. 
For instance, hydrodynamic instabilities~\cite{Meerson2002,lou_hall}, phase transitions associated with changes in system's symmetry~\cite{olafsen_urbach_2005,Pagonabarraga2018} and memory formation~\cite{Jaeger2000,Ihle2017} display a rich phenomenology in a wide variety of materials, including granular fluids, active fluids, magnetic colloids, etc.~\cite{Nagel2019}. 
Most interestingly, phase separation appears to be ubiquitous in granular~\cite{goldhirsch_clustering,argentina_vdw} and active fluids~\cite{Cates_2015,caporusso_dumbbells2d,caporusso_dumbbells3d}, controlled by mechanisms that extend far beyond their equilibrium counterparts.

In granular matter, energy is dissipated at the single grain scale through frictional forces at collisions. 
To prevent dynamical arrest, an external action needs to be applied, so that by constant input of kinetic energy the system goes into stationary, inherently out of equilibrium states. 
In quasi two-dimensional granular layer experiments, energy can be injected, {\sl e.g.} shaking the sample along the third direction, and grains gain energy from collisions, both with the confining walls and neighboring particles~\cite{olafsen_shakers,Melby_2005}. 
In such an experimental set-up, phenomena like crystallization (hexatic~\cite{olafsen_urbach_2005} and cubic~\cite{prevost_coexistence}) and phase separation~\cite{soto_gran_phase_separation}, beyond their equilibrium counterparts, have been realized.

As opposed to granular fluids, where particle translations and rotations are passive, active fluids are characterized by a mechanism that produces systematic translation and/or rotation at the particle level, leading generically to new types of instabilities and phase transitions, such as flocking~\cite{Viseck_1995} and motility-induced phase separation~\cite{Cates_2015,Puglisi_2020}.
Although phase separation has also been observed recently in chiral fluids with active rotations, no generic mechanism has yet been identified~\cite{lintuvuori_cavitation,caprini_bubble}.

Recently, a system of disks with tilted radial blades has been experimentally proposed. 
Disks acquire persistent rotation when subjected to air upflow~\cite{fran_PRR,fran_frontiers,fran_commphys,fran_pof,farhadi_airspinners,workamp_airspinners}. 
Complementarily, experimental systems with similar chiral interactions, ranging from self-spinning grains and robots~\cite{tsai_chira_gran,liu_gran_spinners,yang_exp_PRE,scholz_spinning_robots,li_memory_spinners} to magnetic colloids~\cite{stone_mag_spinners,yan_mag_spinners,snezhko_mag_spinners,soni_cubes,han_mag_spinners,massana_mag_spinners,bililign_chir_defs,stone_mag_spinners} and optically activate micro-rotors~\cite{modin_light_spinners} have been proposed. 
A generic property of transport phenomena in these fluids is that their fluxes inherently have an anti-symmetric component. 
As a consequence, a set of new transport coefficients, labeled as \textit{odd}, emerges~\cite{avron_odd_viscosity}.
The collective behavior of chiral fluids in these systems has been explored theoretically and computationally~\cite{van_zuiden_spatiotemporal,han_fluctuating_2021,reevs_spinners_lanes,liebchen_levis_review,lou_hall,fruchart_odd_viscosity,huang_odd_elasticity_granular,lintuvuori_cavitation,caporusso_edge,caprini_bubble}, but, despite these efforts, the possibility that a chiral fluid undergoes generically a phase separation with specific features due to its oddness remains an open question. 
We show here that phase separation may be generically induced in chiral fluids from the balance between rotational activity and friction, features that are generic in chiral fluids. 

To this end, we introduce a minimal model describing a system of $N$ identical rotating disks with mass $m$, diameter $\sigma$ and moment of inertia $I$. 
The disks are subject to forces and torques coming from three sources: friction forces and torques with the embedding medium (with translational, $\gamma$, and rotational, $\gamma_\theta$, friction coefficients), interparticle forces, and an active torque. 
The system is determined by a Langevin dynamics where the thermal bath, of temperature $T_{\rm{th}}$, is a white noise for both translation, $\boldsymbol{\xi}$, and rotation, $\boldsymbol{\xi}_{\theta}$, with zero mean and time correlations $\langle \boldsymbol{\xi}_i(t) \boldsymbol{\xi}_j(t^{\prime}) \rangle = 2\gamma T_{\rm{th}} \bm{1} \delta_{ij} \delta(t-t^{\prime})$ and $\langle \boldsymbol{\xi}_{\theta,i}(t) \boldsymbol{\xi}_{\theta,i}(t^{\prime}) \rangle = 2\gamma_{\theta} T_{\rm{th}} \bm{1} \delta_{ij} \delta(t-t^{\prime})$. Namely,
\begin{align}
    \label{eq:motion_tr}
    m \frac{{\rm{d}} \bm{v}_i}{{\rm{d}}t} &= -\gamma \ \bm{v}_i + \boldsymbol{\xi}_i + \mathbf{F}_i^\mathrm{n} + \bm{F}_i^{\rm{t}} \ \rm{,} \\
    \label{eq:motion_rot}
    I \frac{{\rm{d}}\bm{\omega}_i}{{\rm{d}} t} &= -\gamma_{\theta} \ \bm{\omega}_i + \boldsymbol{\xi}_{\theta,i} + \boldsymbol{\tau}_i + \boldsymbol{\tau}_0 \ \rm{.}
\end{align}

\begin{figure}[t!]
    \centering
    \includegraphics[width=\columnwidth]{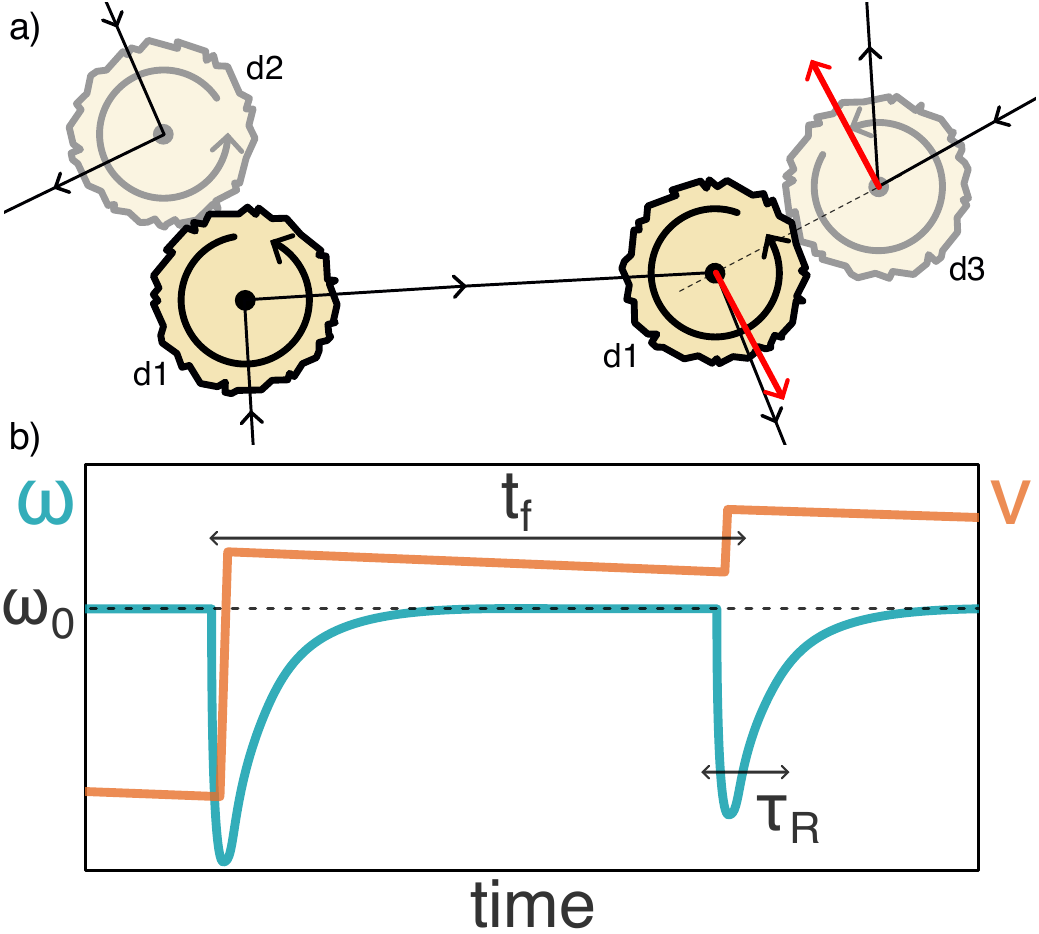}
    \caption{
    (a) A sketch of two consecutive collisions of a disk d1 (solid colors), with corresponding incoming and outgoing trajectories (black lines and arrows). 
    The first collision is with disk d2 (transparent), the second one is with disk d3 (transparent as well). 
    The disks are represented with a rough surface, denoting the action of a tangential friction-like force (symbolized as red arrows in the case of the second collision). 
    (b) Time evolution of disk d1 speed ($v$, orange line) and angular velocity ($\omega$, cyan line), over a time interval that includes the two consecutive collisions. 
    The rotational inertia time scale, $\tau_R$, and the free flight time, $t_f$, are the two relevant timescales. 
    In this case, rotational dynamics is much faster, hence $\tau_R<t_f$, and the disk recovers the intrinsic spinning velocity, $\omega_0$, before colliding again.
    The translational inertial timescale, $\tau_\mathrm{I}$, not displayed here, is associated to friction and is much larger than the other two, since in our system energy loss due to friction in between two collisions is small.
    }
\label{fig:spinners}
\end{figure}

The interactions between disks are controlled by normal and tangential forces. 
The former, $\mathbf{F}_i^\mathrm{n} = -\boldsymbol{\nabla}_i \sum_{j\neq i} U(|\bm{r}_j-\bm{r}_i|)$, is due to a purely repulsive Weeks-Chandler-Anderson (WCA) potential \cite{Weeks_1971}, $U(r)= 8\varepsilon \bigl[ 2(\sigma/r)^{12}-(\sigma/r)^6 \bigr]$, cut at its minimum $r_c=\sigma$. 
The tangential force, $\bm{F}_i^{\rm{t}} = - \sum_{j\neq i} \eta \ \bm{v}_{ij}^{\rm{t}}$, of range $r_{ij}<\sigma$ ($r_{ij}$ being the interparticle distance), is due to friction between particles and characterized by a friction coefficient, $\eta$. 
The relative tangential velocity at the contact point is defined as $\bm{v}_{ij}^{\rm{t}} = \bm{v}_{ij} - (\bm{v}_{ij} \cdot \hat{\bm{n}}_{ij})\hat{\bm{n}}_{ij} - \bm{\omega}_{ij} \times \hat{\bm{n}}_{ij}$, where $\bm{v}_{ij}=\bm{v}_j-\bm{v}_i$, $\bm{\omega}_{ij}=(\bm{\omega}_i+\bm{\omega}_j)\sigma/2$ and $\hat{\bm{n}}_{ij}=(\bm{r}_i-\bm{r}_j)/r_{ij}$. 
The tangential force induces an angular momentum impulse upon particle encounters, due to the torque $\boldsymbol{\tau}_i=-\frac{1}{2}\sum_{j\neq i}({\bm{r}}_{ij} \times \bm{F}_{ij}^{\rm{t}})$. 
Finally, disks are subject to a homogeneous and constant torque $\boldsymbol{\tau}_0$, which generates a persistent rotation at a fixed rate $\omega_0=\tau_0/\gamma_{\theta}$; the source of active rotations, also referred to as spin.

Fig.~\ref{fig:spinners} shows the combined action of tangential friction and active rotation in  disks' collisions, which breaks the time parity when converting rotational into translational kinetic energy. 
In particular, the rotational kinetic energy may drop at collisions, relaxing to $\omega_0$ as imposed by the external forcing after the collision. 
Consequently, the translational kinetic energy shows a sharp increase in collisions, and subsequently decreases because of medium friction. 
Overall, rotational kinetic energy is injected into the system at particle level and transformed into translational kinetic energy upon collisions, breaking detailed balance and driving the system towards an out of equilibrium and chiral state.

\begin{figure}[t!]
    \centering
    \includegraphics[width=\columnwidth]{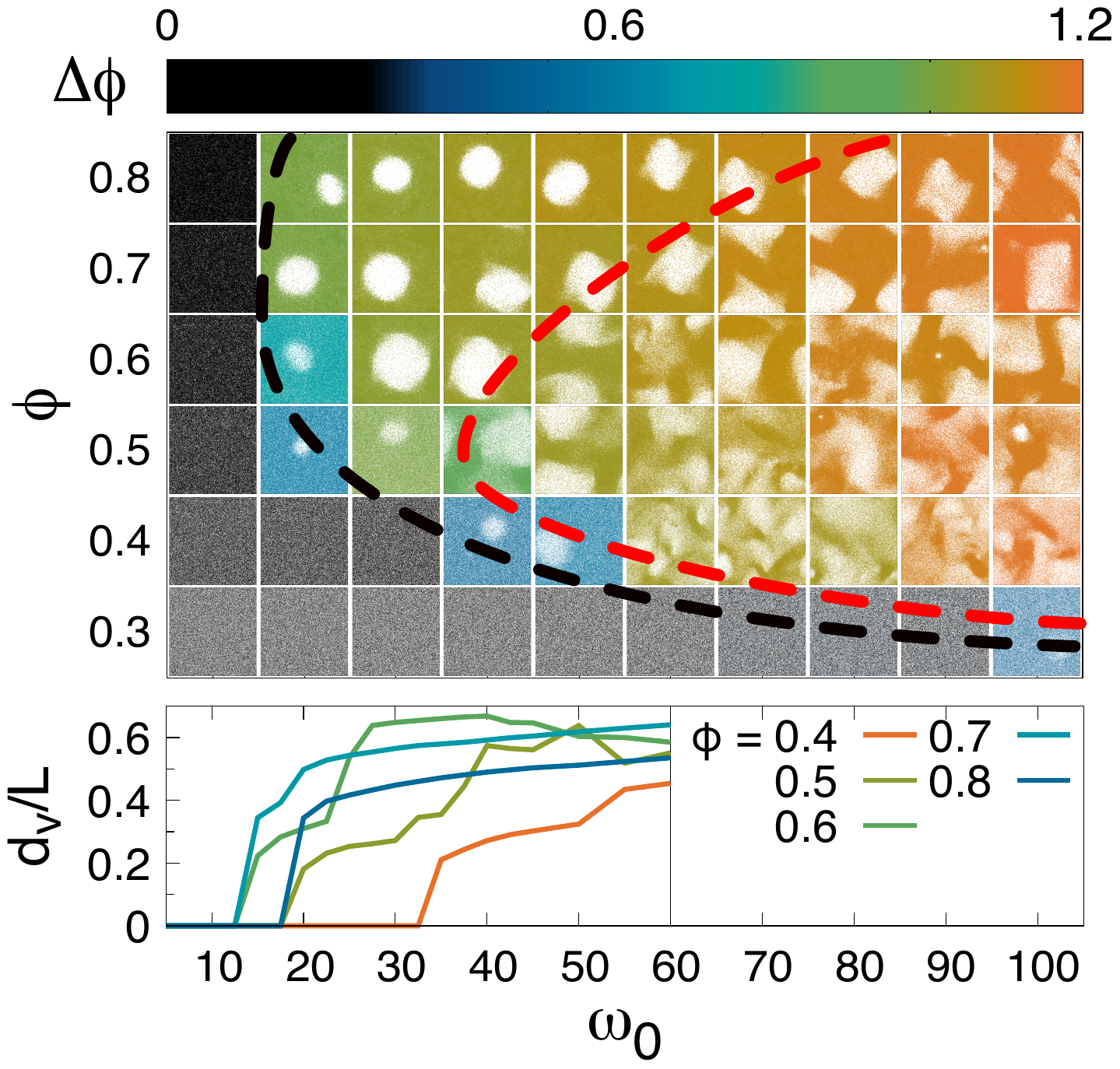}
    \caption{
    A collection of representative configurations, sampled in the late time regime, in the $\omega_0-\phi$ plane. Particles are colored according to the dispersion in local density, $\Delta \phi=\phi_{\rm{loc}}^{\rm{max}}-\phi_{\rm{loc}}^{\rm{min}}$ (white stands for particle-free spaces).
    The system displays phase separation within the region enclosed by the black dashed curve.
    The red dashed curve separates the phase separation region into two. To the left, the phase separation pattern is stationary, with the gaseous phase forming a circular stationary void, surrounded by a dense phase.
    To the right, the pattern becomes turbulent, with the two phases moving and rearranging in time.
    The bottom panel displays the diameter of the gaseous voids for stationary phase separation ($d_V$), as a function of $\omega_0$, for different values of $\phi$.
    }
\label{fig:confs}
\end{figure}

We have performed molecular dynamics (MD) simulations~\cite{LAMMPS} of $N=256^2$ disks following the equations of movement Eqs.~\eqref{eq:motion_tr},~\eqref{eq:motion_rot} in a system of size $L$, and explore the effect of intrinsic spin, $\omega_0$, and surface fraction, $\phi=N \pi \sigma^2/4L^2$ on the collective behavior of the system. 
Unless otherwise stated, $T_{\rm{th}} = 1/100$, $\gamma=1/10$, $\gamma_{\theta}=10$, $\eta=100$, in units of mass $m$, length $\sigma$ and energy $\varepsilon$.
 
Spinners possess translational and rotational inertia, characterized by time scales $\tau_I=m/\gamma$ and $\tau_R=I/\gamma_{\theta}$, respectively. 
We focus on the regime $\tau_R \ll \tau_I$, where chiral effects are strong. 
For an arbitrary rotating disk, the post-collisional spin, $\omega$, relaxes rapidly to $\omega_0$, while the translational velocity,  $v$,  slowly decays due to friction with the surrounding medium (Figure~\ref{fig:spinners} (b)). 

For large enough $\phi$ and $\omega_0$, the system groups into dense regions that leave nearly empty areas. 
Fig.~\ref{fig:confs} (a) shows representative configurations of the system in the late-time regime. 
As it can be seen, the phenomenology of this Rotation Induced Phase Separation (RIPS) is very rich with regard to the changes in the structure of the coexisting phases, the shape of their interface and their dynamics, as $\omega_0$ and $\phi$ are varied.

Increasing $\omega_0$ at constant $\phi$, we identify a value above which the system develops a gaseous void that is surrounded by a dense phase. 
The threshold curve for the phase-separated regime is indicated by the dashed black line in the phase diagram of Fig.~\ref{fig:confs}, characterized by a critical persistence angular velocity, $\omega_{0,c}$.
Above $\omega_{0,c}$, the void increases monotonically until $\omega_0$ reaches a second critical value, $\omega_{0,\rm{ns}}$, where the void is destabilized in favor of an unsteady regime. 
This second transition is indicated by the red dashed line in Fig.~\ref{fig:confs} (a). 
For $\omega > \omega_{0,\rm{ns}}$, the phase-separated pattern shows chaotic dynamics with the voids and the dense phase constantly moving and rearranging.
Movies of the system dynamics in the different regimes are provided in \textit{Appendix A}.
We observed that, regardless of the values of $\omega_0$ and $\phi$, RIPS is suppressed in the absence of large enough translational inertia, see \textit{Appendix B}.

A similar path is followed when $\phi$ increases at constant $\omega_0$, with a critical value of $\phi$ for the appearance of the void. 
As shown in the bottom panel of Fig.~\ref{fig:confs} (b), the typical size of the void is now reentrant for larger $\phi$. 
Since voids reach a maximum typical size, as the system size is increased, RIPS is characterized by a lattice of chiral voids in the stationary phase-separated regime, see Fig.~\ref{fig:large_system} and the movie linked in \textit{Appendix A}.

\begin{figure}[t!]
    \centering
    \includegraphics[width=\columnwidth]{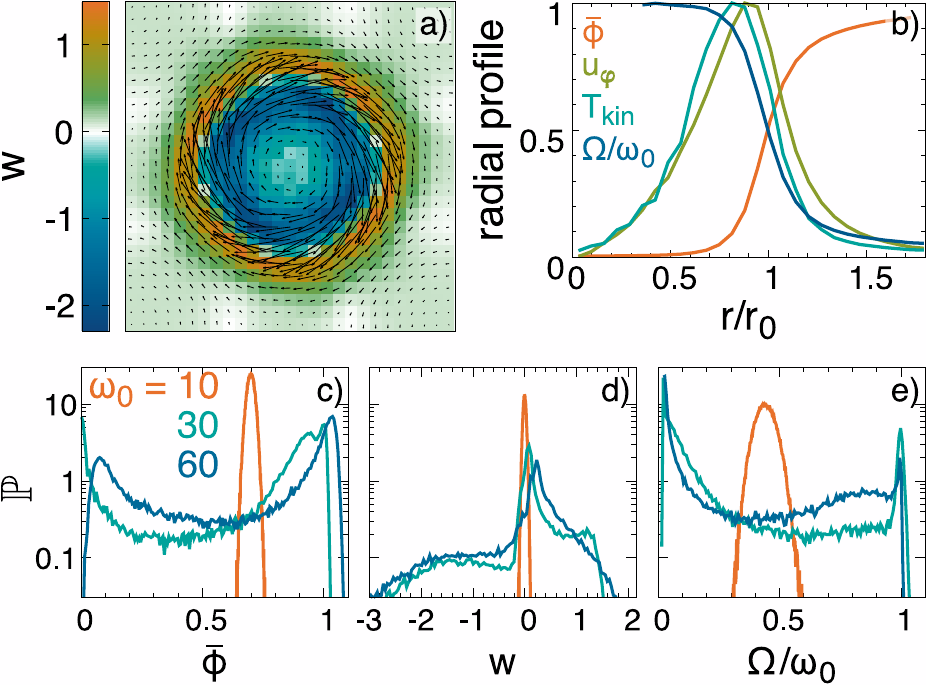}
    \caption{
    (a) Velocity field at $\omega_0=30$ and $\phi=0.700$. 
    Colors represents vorticity $w=\partial_y u_{x}-\partial_x u_{y}$, where $\bm{u}$ is the flow velocity field. 
    (b) Radial profile corresponding to the configuration in (a), computed with respect to the center of the void, of the local density $\bar{\phi}$, the azimuthal component of the flow velocity $u_{\varphi}$, the kinetic temperature $2{T}_{\rm{kin}}/m\equiv(\bm{v}-\bm{u})^2$, and the local average angular velocity field $\Omega$, normalized by $\omega_0$. 
    The values of $u_{\varphi}$ and $T_{\rm{kin}}$ are scaled with respect to their maximum values, $u_{\varphi}^{\rm{M}}\approx 50$ and $T_{\rm{kin}}^{\rm{M}}\approx300$. 
    In the lower panels are shown the probability density functions of (c) local density, (d) local vorticity, (e) local angular velocity, for $\phi=0.700$ and $\omega_0=10,30,80$ in the homogeneous phase, separated phase with a stationary void, non stationary phase, respectively.
    }
\label{fig:void}
\end{figure}

RIPS voids are circle-shaped under the regime of stationary pattern, and there is a counter-spinwise edge current over the interface (i.e., opposite to single-particle intrinsic spin). 
This is shown in the velocity field of Fig.~\ref{fig:void} (a) and the radial profile of the azimuthal (perpendicular to the radial component) flow velocity $u_\varphi$ and translational temperature $T_\mathrm{kin}$ of Fig.~\ref{fig:void} (b).
These currents generate a vorticity pattern, with a negative (counter-spinwise) vorticity inside the void that switches to positive (spinwise) at the interface and then decays in the dense phase. 
Fig.~\ref{fig:void} (b) displays the characteristic pattern of the local spin $\Omega = \langle \omega_i \rangle_{\rm{loc}}$, with $\Omega \simeq \omega_0$ in the interior of the void, where particles are free to move and unlikely to collide. 
$\Omega$ drops in the dense phase, where rotational motion is suppressed by tangential forces over more frequent collisions. 
These features of the phase separated state can be quantified through the probability density function $\mathbb{P}$ of the local surface fraction $\bar{\phi}$, the local vorticity $w$ and the angular velocity field $\Omega$, as shown in Fig.~\ref{fig:void} (c-e), respectively, and compared with the case of a homogeneous system at $\omega_0=10$, where we see no net vorticity and a normal-distributed spin.
In the phase separated regime, we observe a bimodal distribution for $\bar{\phi}$ and $\Omega$. 
The maxima for the bimodal distributions correspond to a dense phase with packing fraction peaked close to 1 and spin peaked around zero and a loose phase with packing fraction peaked close to zero and spin peaked around $\omega_0$. 
The edge currents produce the concurrent appearance of net positive and negative local vorticity.

\begin{figure}[t!] 
    \centering
    \includegraphics[width=\columnwidth]{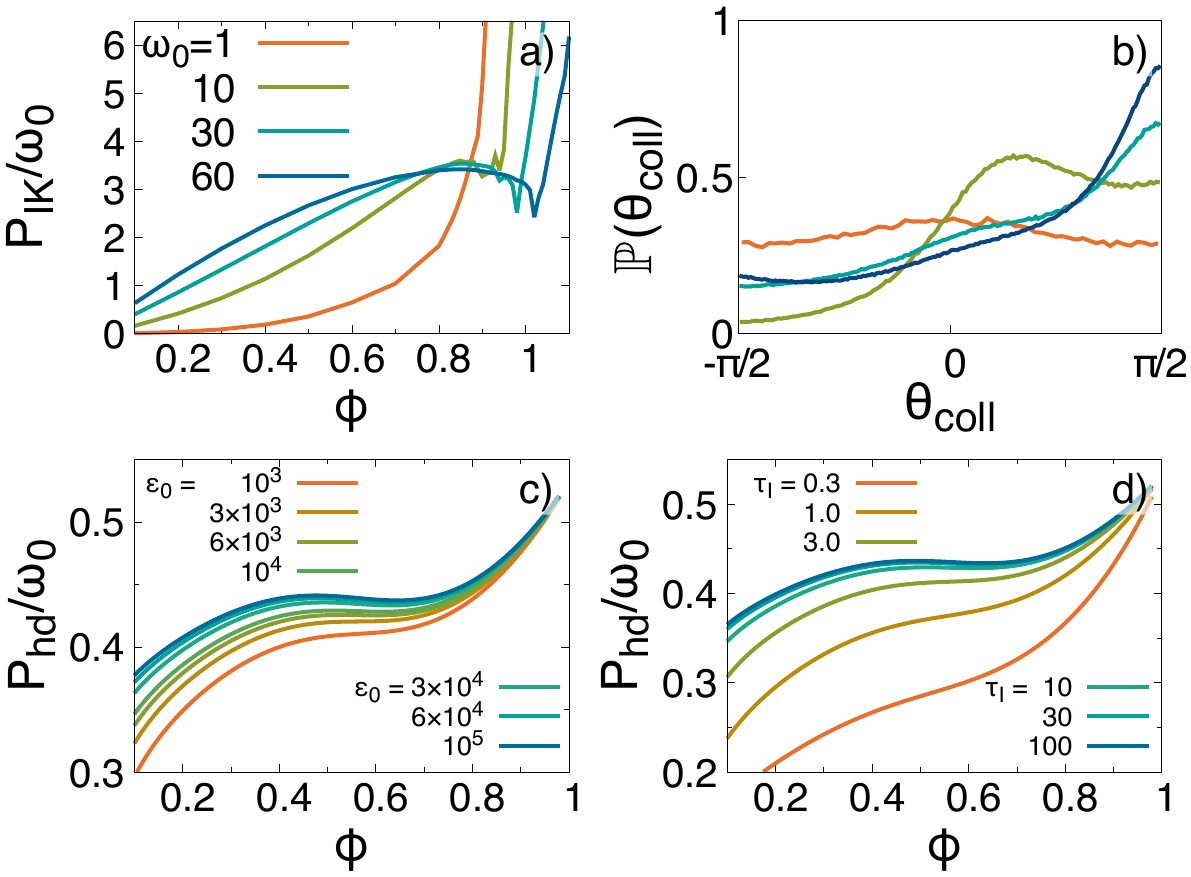}
    \caption{
    (a) Numerical pressure $P_{\rm{IK}}$ for different values of $\omega_0$.
    (b) Probability density function $\mathbb{P}$ of the collisional angle, measured from simulations, for $\phi=0.700$ and the same values of $\omega_0$ as in (a).
    Hard disk pressure, $P_{\rm{hd}}$, computed within the kinetic model for $\alpha=0.1$, $\tau_R=1.0$, $\nu=3$, for (c) $\tau_I=10$ and increasing values of $\varepsilon_0$ and (d) $\varepsilon_0=10^4$ and increasing $\tau_I$.
    Values of numerical and effective pressure are scaled by $\omega_0$ for graphical convenience.
    }
\label{fig:model}
\end{figure}

To understand the origin of RIPS, we compute the Irving-Kirkwood pressure~\cite{ik_pressure} $P_{\rm{IK}}=\frac{1}{V} \sum_{i=1}^N \bigl[ m v_i^2+\bm{r}_i \cdot (\bm{F}_i^t+\bm{F}_i^n) \bigr]$ from a small system that remains homogeneous for all values of $\omega_0$ and $\phi$. 
Fig.~\ref{fig:model} (a) shows that $P_{\rm{IK}}$ develops, at high enough $\omega_0$, a non-monotonic behavior with a regime of negative compressibility. 
This can lead to an instability that eventually drives phase separation, as also observed in heated granular fluids~\cite{soto_gran_phase_separation}.

In order to gain insight into the instability mechanism, we propose a kinetic model for 2D hard disks, to compare the hard disk pressure, $P_{\rm{hd}}=4/(\pi \sigma^2) \phi T (1+2\phi \chi_{\rm{hd}})$~\cite{hd_pressure}, with MD results.
As has been proposed in shaken granular gases~\cite{soto_gran_phase_separation}, we use an equilibrium pair distribution function at contact, $\chi_{\rm{hd}}=(1-7\phi/16)/(1-\phi)^2$, to derive $T$ as an effective temperature resulting from the balance of energy exchange between colliding particles and between particles and the substrate.

Between collisions, particles lose translation energy due to the friction with the substrate at a rate controlled by the particles' inertia, which can be expressed as
\begin{equation}
\label{eq:DEk_fric_ave}
    \langle \Delta E_k^{\rm{fric}} \rangle = \langle{\frac{1}{2} \int_0^{t_{\rm{f}}} {\rm{d}}t \ \gamma \ v_{\rm{out}}^2 {\rm{e}}^{-2t/\tau_I}}\rangle 
    = \frac{\langle t_f \rangle}{2 \langle t_f \rangle +\tau_I} T \ {\rm{,}}
\end{equation}
where $v_{\rm{out}}$ is the outgoing velocity after a collision, and in the averages we have assumed a Maxwell-Boltzmann distribution for the velocities and an exponential free flight time distribution, with $\langle t_f \rangle=\sigma \sqrt{m\pi}/(8 \phi \chi_{\rm{hd}} T^{1/2})$~\cite{luding_gran}.

The gain of translational kinetic energy of a pair of colliding particles, as prescribed by Eqs.~\eqref{eq:motion_tr},~\eqref{eq:motion_rot}, reads
\begin{equation}
\label{eq:DEk_coll}
    \Delta E_k^{\rm{c}} = m \alpha \bigl[ \Delta v^{\rm{t}}((\alpha-1)\Delta v^{\rm{t}} + (1-2\alpha) \sigma \omega) + \alpha (\sigma \omega)^2\bigr]{\rm{,}}
\end{equation}
where $\Delta v^{\rm{t}}=\Delta v \ \sin{\theta}$ is the  incoming relative tangential velocity and $\theta$ the collision angle (See Fig.~\ref{fig:coll_diagram} and \textit{Appendix C} for the definition of $\theta$ and the derivation of Eq.~\eqref{eq:DEk_coll}).
We have defined $\alpha=\eta \Delta t/m$, with $\Delta t$ the typical collision time and $\omega$ the average incoming angular velocities of the pair.
As illustrated in Fig.~\ref{fig:spinners}, $\omega$ evolves dynamically prior to the collision, as $\omega = 2\omega_0 \bigl[ 1-{\rm{e}}^{-t_f/\tau_R} \bigr]$, assuming that $\omega$ drops to zero at all collisions. 
Hence, the average energy exchange at collisions reads
\begin{equation}
\label{eq:DEk_coll_ave}
    \begin{split}
    \langle \Delta E_k^{\rm{c}} \rangle &= \frac{8\alpha^2\varepsilon_0 \langle t_f \rangle^2}{(\tau_R+\langle t_f \rangle)(\tau_R+2\langle t_f \rangle)} + 2\alpha(\alpha-1) \ T \langle \sin^2{\theta} \rangle_{\theta} \\
    &+ \alpha(1-2\alpha) \sqrt{2\pi \varepsilon_0 T}  \biggl( \frac{\langle t_f \rangle}{\tau_R+\langle t_f \rangle} \biggr) \langle \sin{\theta} \rangle_{\theta}
    \end{split}
\end{equation}
where $\varepsilon_0=m \sigma^2\omega_0^2/4$ stands for the particle rotational energy input, and $\langle \rangle_{\theta}$ for an average over collision angles.

As interactions are non-central, inhomogeneous states develop chiral flows with a relevant impact on the collision angle distribution, $f(\theta)$. Fig.~\ref{fig:void} (a) shows that particles flowing along a void edge are more likely to undergo grazing collisions with those in the bulk, with a preferred angle around $\theta=\pi/2$.
This effect is captured by 
\begin{equation}
\label{eq:theta_dist}
    f(\theta) = \frac{\left(\frac{\phi}{1-\phi}\right)^{\nu} + 2\pi \ \delta(\theta-\frac{\pi}{2})}{\pi \left[1+\left(\frac{\phi}{1-\phi}\right)^{\nu}\right]} \ ,
\end{equation}
with $\theta \in [-\pi/2,\pi/2]$; and backed up by MD as shown in  Fig.~\ref{fig:model} (b), which displays a peak at $\pi/2$ as $\omega_0$ increases.

A steady state is achieved when Eqs.~\eqref{eq:DEk_fric_ave} and~\eqref{eq:DEk_coll_ave} balance each other, determining the translational kinetic temperature, $T$ (the expression can be found in \textit{Appendix C}), which results a monotonically decreasing function of $\phi$.
Interestingly, the corresponding hard disk pressure displays a van der Waals loop. 

In agreement with MD results, Fig.~\ref{fig:model} (c,d) show a transition from an increasing pressure to a non-monotonic one with respect to $\varepsilon_0$ and/or $\tau_I$. 
This can be interpreted as a competition among the relevant timescales at play.
Finite rotational inertia makes the rotational to translational kinetic energy exchange increasingly concealed by contact friction for increasing density. 
As $\tau_R > t_f$ at high density, active rotation is not fully recovered in between collisions, and hence the fluid becomes less chiral. 
This explains why the temperature decreases with density, thus yielding a decreasing pressure. 
$\tau_I$ decreases upon increasing friction with the embedding medium, causing the energy gained in collisions to be quickly dissipated, cooling the system at low density.
The last crucial ingredient for the appearance of a van der Waals loop is the form of $f(\theta)$ in Eq.~\eqref{eq:theta_dist}, which accounts for the effect of chiral currents, and modulates the dependence on density of the energy contribution in Eq.~\eqref{eq:DEk_coll_ave}. 
The relevance of this term is evident from the fact that the van der Waals loop is present only for $\nu\geq2$.
For lower values of $\nu$, the decrease in $T$ with $\phi$ is not fast enough to allow a negative pressure slope.
Further details on the dependence of the pressure on $\nu$ are reported in \textit{Appendix D}.
This means that when a density gradient is created in the system due to spontaneous fluctuations, the emergence of the circulating currents favors an extra energy input through the selection of preferred collision angles, corresponding to grazing collisions, which triggers the mechanical instability responsible for phase separation.

In summary, we have shown that the RIPS develops as a result of the pressure that emerges from the competition between active rotation and interparticle friction.
Rotation transfers linear momentum more efficiently at low densities, while contact friction inhibits particles' persistent rotation in the denser region. 
As a result, a positive feedback mechanism is induced, which increases pressure in low-density regions and reduces pressure in high-density regions. 
Consequently, particles are expelled out of the void region towards the denser region.
This is reflected in the emergence of a Van der Waals loop, as shown in Fig.~\ref{fig:model}. 
We observe that RIPS is characterized by voids of finite size, as they grow until they become so rarefied that particle encounters are not frequent enough, angular to linear momentum transfer becomes inefficient, and the system reaches a stationary state.
A more detailed description of the mechanical balance between the two coexisting phases, including the odd components of the stress that generates the circulating currents~\cite{lou_hall}, is required to predict the void size.

The proposed model has the minimal features that explain the emerging mechanical instability leading to the phase separation observed for inertial spinners. 
A similar phenomenon has been observed in systems that interact through noncentral forces and generate chiral flows~\cite{lintuvuori_cavitation,caprini_bubble}. 
Therefore, RIPS is a general phenomenon associated to a large class of systems that share a few simple features, such as an intrinsic spin, short range tangent forces, and translational inertia.

\begin{acknowledgments}
The authors are thankful to Prof. R. Soto for fruitful discussion. IP acknowledges the support from Ministerio de Ciencia, Innovaci\'on y Universidades MCIU/AEI/FEDER for financial support under grant agreement PID2021-126570NB-100 AEI/FEDER-EU, and Generalitat de Catalunya for financial support under Program Icrea Acad\`emia and project 2021SGR-673. FVR acknowledges support from Ministerio de Ciencia, Innovaci\'on y Universidades through Agencia Estatal de Investigación (AEI) project no. PID2020-116567GB-C22 and Estancias de Movilidad (Modalidad A) fellowship (ref. PRX21/00490).
\end{acknowledgments}

\bibliography{refs}

\appendix

\newpage
\onecolumngrid
\vspace{\columnsep}
\begin{center}
\rule[3pt]{0.4\textwidth}{0.4pt}
\textbf{\large{ \ End Matter \ }}
\rule[3pt]{0.4\textwidth}{0.4pt}
\end{center}
\vspace{\columnsep}
\twocolumngrid

\setcounter{equation}{0}
\setcounter{figure}{0}
\setcounter{table}{0}
\renewcommand{\theequation}{A\arabic{equation}}
\renewcommand{\thefigure}{A\arabic{figure}}

\section{Appendix A. Description of the movies and finite-size effects}

The supplemental movie ``movie1.mov"~\cite{movies} shows the late time dynamics of the system at global surface fraction $\phi=0.700$, for three different values of the persistent angular velocity, namely $\omega_0=10$, in the homogeneous phase, $\omega_0=30$, in the phase separated stationary phase, and $\omega_0=80$, in the phase separated non stationary phase (see the phase diagram in Fig.~\ref{fig:confs} of the main text).
Particles are colored by their instantaneous speed.

\begin{figure}[h!]
    \centering
    \includegraphics[width=\columnwidth]{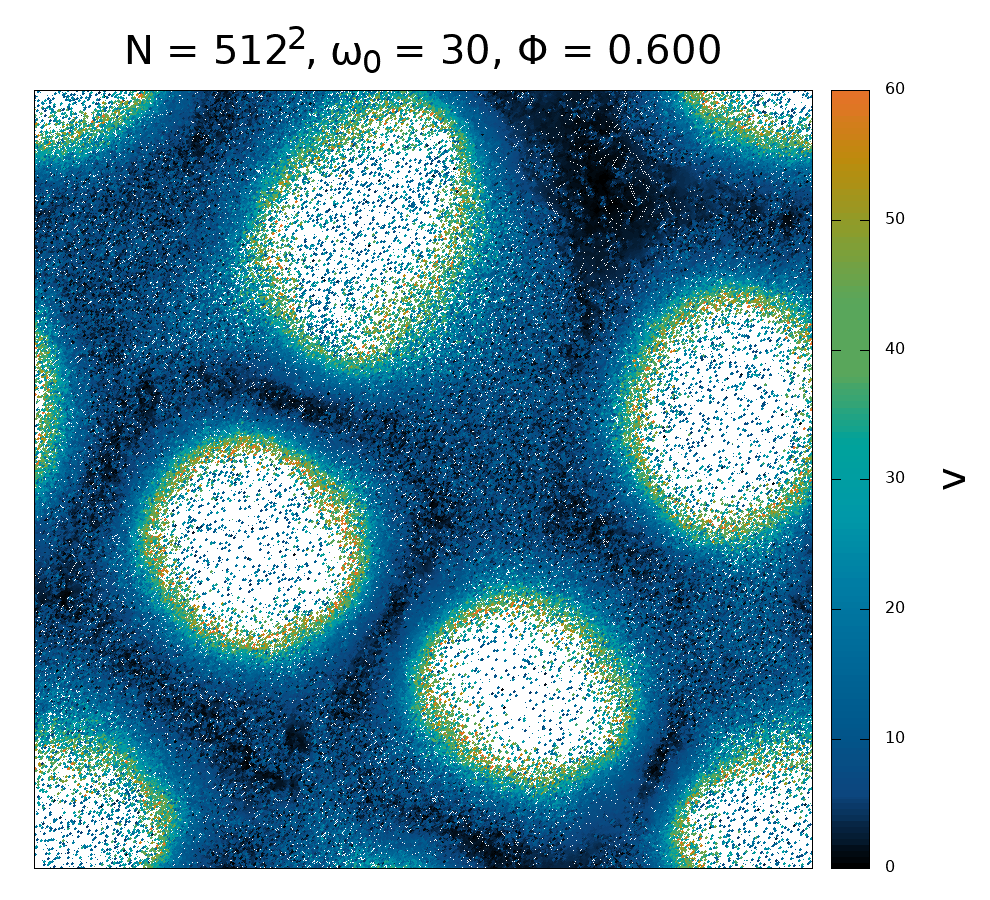}
    \caption{
    Configuration of a system of $N=512^2$ spinners, with $\omega_0=30$ and $\phi=0.600$.
    Colors represent the particles' instantaneous velocities, showing edge currents at the voids' interface.
    }
\label{fig:large_system}
\end{figure}
    
As mentioned in the main text, the typical size of the voids in the RIPS regime is finite, reaching a characteristic maximum value for each system configuration. As shown in Fig.~\ref{fig:large_system}, this implies that, simulating a larger system, compared to the one seen in the main text, it shows that RIPS is characterized by arrested phase separation, forming an array of voids of approximately the same size.
The supplemental movie ``movie2.mov"~\cite{movies} shows that this configuration is stationary and the voids do not coarsen, while generating the same edge current pattern as in the smaller system.

\setcounter{equation}{0}
\setcounter{figure}{0}
\setcounter{table}{0}
\renewcommand{\theequation}{B\arabic{equation}}
\renewcommand{\thefigure}{B\arabic{figure}}

\section{Appendix B. Suppression of phase separation for small inertia}

\begin{figure}[h!]
    \centering
    \includegraphics[width=\columnwidth]{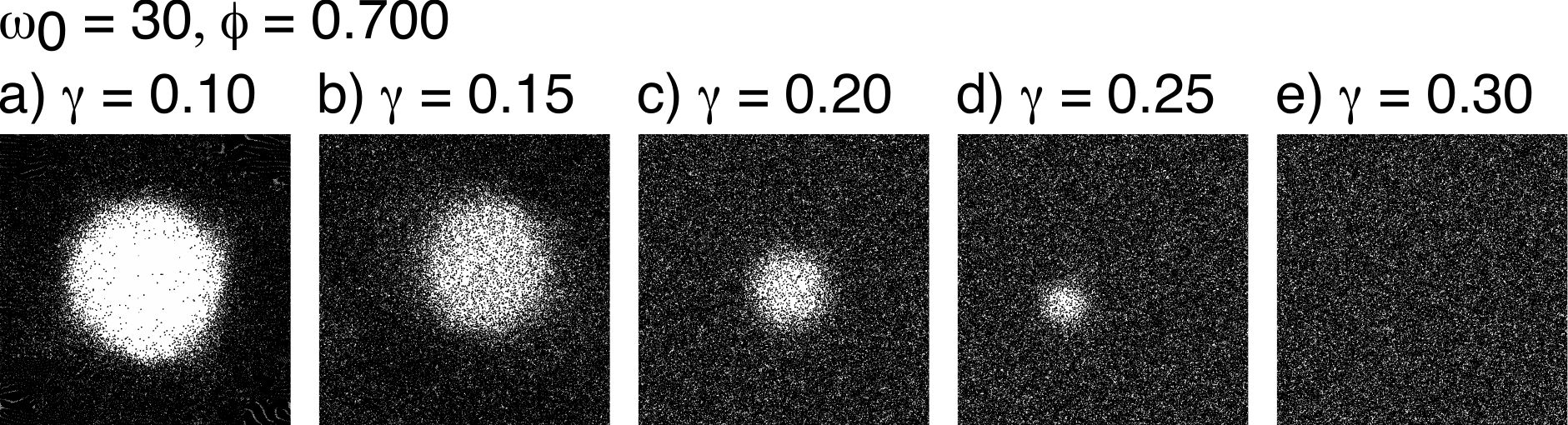}
    \caption{
    Representative configurations of the system, sampled within the late time stationary regime, $\omega_0=30, \phi=0.700$ and increasing values of $\gamma$.
    Starting from (a) $\gamma=0.1$, the value used for all the cases shown in the main text, and increasing $\gamma$, we see the phase separation progressively re-enter (b-d), until it is suppressed at around $\gamma=0.30$ (e).
    }
\label{fig:ps_suppression}
\end{figure}

As  discussed in the main text, phase separation is only observed  in the regime of large particles' translational inertia, $\tau_I$.
 Fig.~\ref{fig:ps_suppression} displays  a set of  configurations with decreasing $\tau_I$ (increasing translational damping coefficient $\gamma$), showing how the void  continuously shrinks  and  phase separation disappears. 
As we showed with our effective model for the hydrostatic pressure, this behavior stems from a balance of energy injected in the system at collisions between particles and lost through friction, controlled by two relevant timescales: the average free flight, $t_f$, enslaved to density, and $\tau_I$.
Fig.~\ref{fig:ps_suppression} shows that $\tau_I$ directly controls the size of the void in the phase separated state.

\setcounter{equation}{0}
\setcounter{figure}{0}
\setcounter{table}{0}
\renewcommand{\theequation}{C\arabic{equation}}
\renewcommand{\thefigure}{C\arabic{figure}}

\section{Appendix C. Energy exchange at collisions}

\begin{table*}
    \caption{
    \label{tab:T_coeffs}
    Coefficients of the polynomial expression for $x=T^2$, as it results in the kinetic model.
    }
    \begin{ruledtabular}
    \begin{tabular}{ c l }
        $x^0$ & $4\alpha^2 \varepsilon_0 (m \pi)^{3/2} \sigma^3$ \\
        $x^1$ & $16 \alpha^2 \varepsilon_0 m \pi \sigma^2 \phi \chi \tau_I + \sqrt{2\pi} \ (m \pi)^{3/2} \sigma^3 \alpha(1-2\alpha) \sqrt{\varepsilon_0} \ \langle \sin{\theta} \rangle_{\theta}$ \\
        $x^2$ & $4 \alpha \sqrt{2\pi} \ m \pi \sigma^2 \phi \chi (1-2\alpha) (\tau_I +\tau_R) \sqrt{\varepsilon_0} \ \langle \sin{\theta} \rangle_{\theta} + 2\alpha(\alpha-1) (m \pi)^{3/2} \sigma^3 \ \langle \sin^2{\theta} \rangle_{\theta} - (m \pi)^{3/2} \sigma^3$ \\
        $x^3$ & $16\pi \sqrt{2m} \ \sigma \alpha \phi^2 \chi^2 \tau_R \tau_I \sqrt{\varepsilon_0} \ \langle \sin{\theta} \rangle_{\theta} + 16 m\pi \sigma^2 \alpha(\alpha-1)(\tau_I+3\tau_R) \phi \chi \ \langle \sin^2{\theta} \rangle_{\theta} - 6 m \pi \sigma^2 \phi \chi \tau_R$ \\
        $x^4$ & $32 \sigma \sqrt{m \pi} \ \alpha(\alpha-1) \phi^2 \chi^2 \tau_R (3\tau_I + 2 \tau_R) \ \langle \sin^2{\theta} \rangle_{\theta} -16\sigma \sqrt{m \pi} \ \alpha \ \phi^2 \chi^2 \tau_R^2$ \\
        $x^5$ & $256 \alpha(\alpha-1) \phi^3 \chi^3 \tau_R^2 \tau_I \ \langle \sin^2{\theta} \rangle_{\theta}$
    \end{tabular}
    \end{ruledtabular}
\end{table*}

Due to the tangential force acting between the particles at collision, a conversion of rotational kinetic energy into translational kinetic energy takes place when two particles collide, resulting in a net translational kinetic energy gain that depends on the collision details.

\begin{figure}[h!]
    \centering
    \includegraphics[width=\columnwidth]{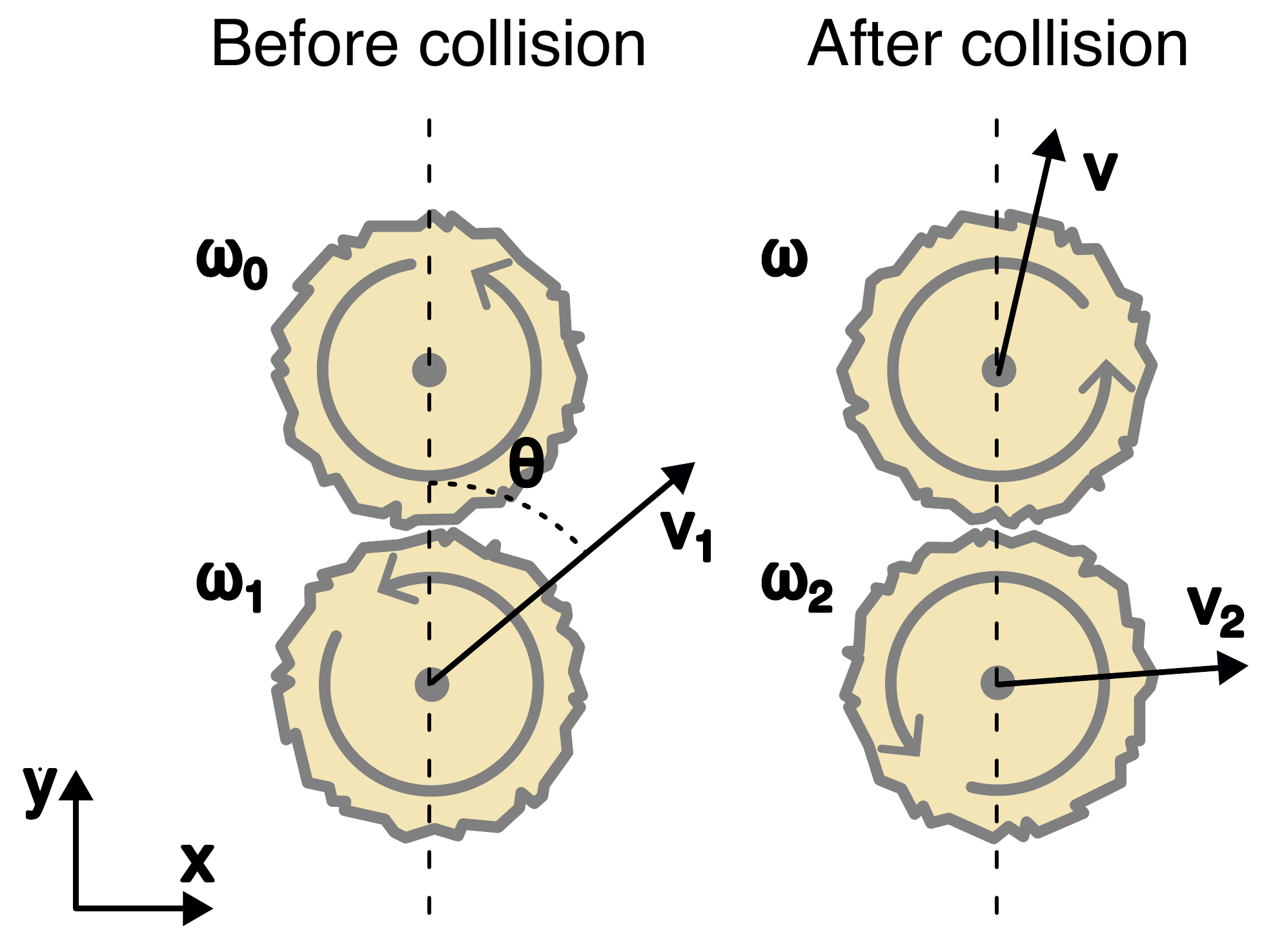}
    \caption{
    Diagram of a general collision between two particles, in the reference frame of the top particle.
    The two particles collide at a generic angle $\theta$, with incoming angular velocities $\omega_0$ and $\omega_1$.
    After the collision, both the translational and rotational velocities change, according to the interaction between the two particles, which is an overlap between a repulsive WCA potential and the transversal force, as described the main text.
    }
\label{fig:coll_diagram}
\end{figure}

As shown in Fig.~\ref{fig:coll_diagram}, the transversal force acting along the $x$-axis for the diagram of a generic collision between two particles can be written as
\begin{equation}
\label{eq:one_coll_tanforce}
    F^{\rm{t}}_x = -\eta \bigl[ v_{1,x}-R(\omega_0+\omega_1) \bigr] \ \rm{.}
\end{equation}
In this geometry, the interaction along the $y$-axis is conservative, thus it does not produce energy variations.

Outgoing velocities and angular velocities are therefore
\begin{equation}
\label{eq:out_vels}
    \begin{split}
        v_{2,x}-v_{1,x} &= \int {\rm{d}}t \frac{F^{\rm{t}}_x}{m} \simeq -\frac{\eta \Delta t}{m} \bigl[ v_{1,x}-\frac{\sigma}{2}(\omega_0+\omega_1) \bigr] \ \rm{,} \\
        v_x&= -\int {\rm{d}}t \frac{F^{\rm{t}}_x}{m} \simeq \frac{\eta \Delta t}{m} \bigl[ v_{1,x}-\frac{\sigma}{2}(\omega_0+\omega_1) \bigr] \ \rm{,} \\
        \omega_2 - \omega_1 &= \int {\rm{d}}t \frac{\tau}{I} \simeq \frac{2\mu \Delta t}{Km\sigma} \bigl[ v_{1,x}-\frac{\sigma}{2}(\omega_0+\omega_1) \bigr] \ \rm{,} \\
        \omega - \omega_0 &= \int {\rm{d}}t \frac{\tau}{I} \simeq \frac{2\mu \Delta t}{Km\sigma} \bigl[ v_{1,x}-\frac{\sigma}{2}(\omega_0+\omega_1) \bigr] \ \rm{,}
    \end{split}
\end{equation}
where $K$ is related to the particles' moment of inertia, $I=K m \sigma^2/4$, and the integral is intended over the duration of the collision $\Delta t$.
The integral is evaluated assuming a very short collision duration, such that the force can be considered constant. One can then compute the outgoing particle velocities, from which Eq.~\eqref{eq:DEk_coll} in the main text is derived.

When the two effective energy contributions in Eqs.~\eqref{eq:DEk_fric_ave} and~\eqref{eq:DEk_coll_ave} balance each other, the system reaches a stationary effective temperature.
The balance condition can be cast in the form of a $5$-th order polynomial equation in the variable $x=T^2$, with coefficients reported in Table~\ref{tab:T_coeffs}.

This equation is solved numerically using the Newton-Raphson method.
We observe that only one positive solution is present, which represents the stationary temperature of the kinetic model.

\setcounter{equation}{0}
\setcounter{figure}{0}
\setcounter{table}{0}
\renewcommand{\theequation}{D\arabic{equation}}
\renewcommand{\thefigure}{D\arabic{figure}}

\section{Appendix D. Impact of oddness in the distribution of collision angles}

\begin{figure}[b!]
    \centering
    \includegraphics[width=0.7\columnwidth]{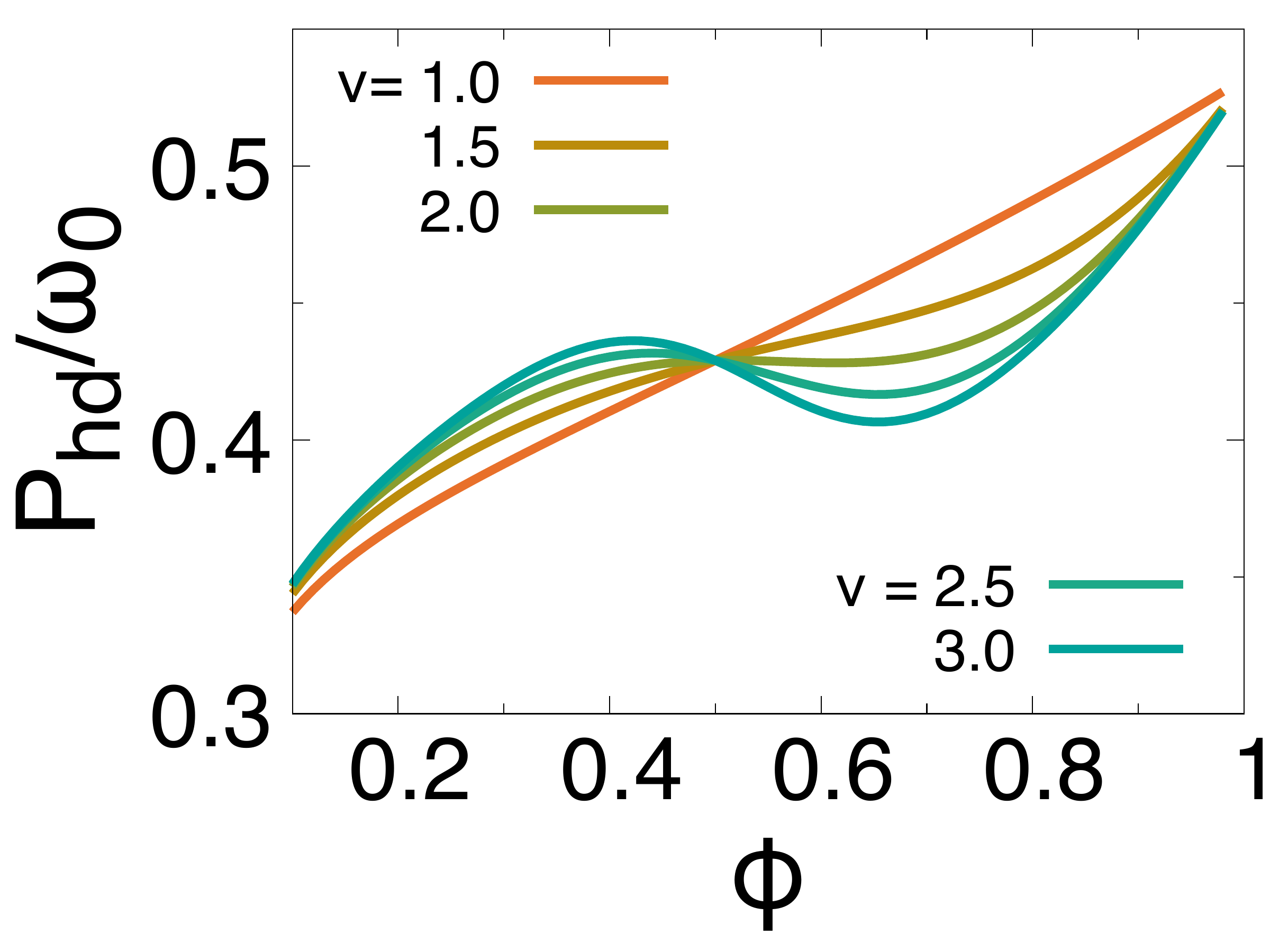}
    \caption{
    Hard disk pressure, as computed from the kinetic model, for $\varepsilon_0 = 10^4$, $\alpha = 0.1$, $\tau_R = 1.0$, $\tau_I = 10$, and different values of $\nu$, which is a parameter in the distribution of collision angles~\eqref{eq:theta_dist}.
    }
\label{fig:odd_coll_pressure}
\end{figure}

Within the developed kinetic model, finite rotational inertia implies rotational to translational kinetic energy conversion suppression at high density, leading to decreasing temperature with density.
However, this condition is not sufficient to develop negative compressibility regimes. 
As shown in the main text, odd interactions are a key feature that enhances energy gain at low density in presence of interfacial currents.
This is shown in Fig.~\ref{fig:odd_coll_pressure}, where the hard disk pressure is displayed for different values of the parameter $\nu$, which represents the degree of oddness in the collision angle distribution employed (see Eq.~\eqref{eq:theta_dist} in the main text).
While the temperature is always a decreasing function of density for the parameters used, the pressure is monotonically increasing for $\nu=1$, and it develops a van der Waals loop for $\nu \ge 2$.

\end{document}